\documentclass[aps,pre,twocolumn,groupedaddress,showpacs]{revtex4-1}

\usepackage[latin1]{inputenc}
\usepackage{amsmath,amssymb,amsfonts} 
\usepackage{graphicx}
\usepackage{psfrag}
\usepackage{color}

\begin{document}

\title{Viscoelasticity of reversibly crosslinked networks of
  semiflexible polymers} \author{Jan Plagge}\altaffiliation{current
  address: Deutsches Institut f\"ur Kautschuktechnologie, e. V.,
  Eupener Str. 33, 30519 Hannover}\author{Andreas
  Fischer}\author{Claus Heussinger} \affiliation{Institute for
  Theoretical Physics, Georg-August University of G\"ottingen,
  Friedrich-Hund Platz 1, 37077 G\"ottingen, Germany}

\begin{abstract}
  We present a theoretical framework for the linear and nonlinear
  visco-elastic properties of reversibly crosslinked networks of
  semiflexible polymers. In contrast to affine models where network
  strain couples to the polymer end-to-end distance, in our model
  strain rather serves to locally distort the network structure. This
  induces bending modes in the polymer filaments, the properties of
  wich are slaved to the surrounding network structure. Specifically,
  we investigate the frequency-dependent linear rheology, in
  particular in combination with crosslink binding/unbinding
  processes. We also develop schematic extensions to describe the
  nonlinear response during creep measurements as well as during
  constant strain-rate ramps.
\end{abstract}


% 83.10.-y 	Rheology Fundamentals and theoretical
% 83.60.Bc 	Linear viscoelasticity
% 83.80.Lz 	Physiological materials (e.g. blood, collagen, etc.)
% 87.16.dm 	Subcellular structure and processes, Mechanical properties and rheology
% 87.16.Ka 	Filaments, microtubules, their networks, and
% supramolecular assemblies
% 87.16.Ln 	Cytoskeleton

 \maketitle

\section{Introduction}

The cytoskeleton is a visco-elastic material with many interesting
mechanical behaviors. From a theoretical point of view these systems
are viewed as networks of reversibly (or permanently) crosslinked
semiflexible
polymers~\cite{bausch06,RevModPhys.86.995,kroy06:_elast}. Over the
years many theoretical works have discovered and partially explained
different regimes, where certain components of the networks dominate
the mechanical
response~\cite{gittesPRE1998,frey98,PhysRevE.63.031502,jones91:_elast_rigid_networ,WolffNJP2010,heu06floppy}. Simulations
on simplified model systems provide a helpful alternative approach to
study the pertinent problems~
\cite{C1SM05022B,Wilhelm2003,Head2003,B815892D,PhysRevLett.112.094303,C0SM00496K}.

Most notably the affine approach, relying on the nonlinear polymer
force-extension relation~\cite{mac95,Storm2005}, has allowed to
rationalize many of the diverse experimental findings. Recent advances
include the glassy wormlike chain model~\cite{kroyNJP2007}, effective
medium
theories~\cite{das12:_redun_cooper_mechan_compos_cross_filam_networ,PhysRevE.87.042601,PhysRevE.85.021801}
as well as models that use analogies with rigidity
percolation~\cite{c.11:_critic_isost}, the jamming transition in dense
particle packings~\cite{LiuNagelSaarloosWyartREVIEW2010} and its
concept of "soft modes". This latter
analogy~\cite{heussingerPRE2007,heussingerEPJE2007} is based on the
fact that densly packed hard particles prefer to rotate around --
instead of press into each other. After all, hard particles are
``incompressible''. Similarly, semiflexible polymers are nearly
inextensible, and under deformation they prefer to deform
perpendicular to the polymer axis -- what is commonly understood as
\emph{bending}.

Here, we present a theoretical framework that is entirely constructed
on the basis of these bending deformations. The force-extension
relation does not play a role for the linear response of the
network. The theory is based on results~\cite{heussingerPRE2007} on
the static linear elasticity. The key achievement of the present work
is that it generalizes these results to finite frequencies, allowing
to calculate the linear elastic and viscous moduli over the whole
frequency regime relevant for standard rheological experiments.

The manuscript is structured as follows: first a brief review of the
static modulus is given (section \ref{sec:revi-stat-modul}). Then
(section \ref{sec:finite-frequency}) the model is generalized to
finite frequencies. In section \ref{sec:finite-crossl-lifet}
low-frequency crosslink binding processes are considered; and finally
(section \ref{sec:nonlinear-response}) we discuss possible nonlinear
rheological effects presenting schematic extensions of the linear
model.

\section{Review: static modulus}\label{sec:revi-stat-modul}

We will consider the properties of a test filament crosslinked into a
network of other filaments.  The filament is described in terms of the worm-like chain
model. In ``weakly-bending'' approximation the bending energy of the
filament can be written as
\begin{eqnarray}\label{eq:hb}
  H_b &=&\frac{\kappa_b}{2}\int_0^L\left(\frac{\partial^2 y}{\partial
      s^2}\right)^2ds
\end{eqnarray}
where $\kappa_b$ is the filament bending stiffness and $y(s)$ is the
transverse deflection of the filament from its (straight) reference
configuration at $y_0(s)=0$. In these expressions $s$ is the
arclength, $s=[0,L]$, and $L$ is the length of the filament.

The effect of the surrounding network is to confine the test filament
to a tube-like region in space.  In this way the actual network is
substituted by an effective potential that acts on the test
filament. A convenient potential is the harmonic tube
\begin{eqnarray}\label{eq:tube}
V = \frac{1}{2}\int_0^L k(s) (y(s)-\bar y(s))^2 ds\,,
\end{eqnarray}
where $k(s)$ is the strength of the confinement and
%$y(s)$ is the transverse deflection of the filament from its (straight)
%reference configuration at $y_0(s)=0$.
$\bar y(s)$ is the tube center, which may or may not be different from
the reference configuration of the filament.
%In these expressions $s$ is the arclength, $s=[0,L]$, and $L$ is the
%length of the filament

%strain couples to tube centerline not to end-distance
A key assumption in our model is that the tube depends on network
strain $\gamma$. In particular, we will assume that the tube
centerline follows the strain, 
\begin{eqnarray}\label{eq:bary}
\bar y(s,\gamma) = \beta(s)\gamma L\,,
\end{eqnarray}
with a shape function $\beta(s)$ that is slaved to the local network
structure. The occurence of the filament length $L$ signifies its role
as non-affinity length, up to which network response is non-affine and
determined by local structural features. In fact, such behavior has
been observed in the simulations of
Ref.~\cite{PhysRevLett.108.078102}. One can derive such a scaling from
the assumption of affine motion for the filament centers of
mass~\cite{heussingerPRE2007}.

The physical picture of strain-induced local deformations is thus,
that the preferred location (the tube) of a polymer changes -- and not
primarily the polymer itself. This is the key difference to many
previous works that assume strain to lead to a change in end-to-end
distance of the polymers. The rheology in these models then is a
direct consequence of the force-extension relation of the single
polymer. By way of contrast, in our approach, the force-extension
relation plays no role at all (for the linear response), and the
polymers can be taken to be completely inextensible.

In fact, one can show~\cite{heussingerPRE2007} that tube deformations
leave the end-to-end distance (to linear order) unchanged, as long as
one takes $\beta_i\equiv \beta(s_i)=-\cot\theta_i$ at crosslink
position $s_i$, where $\theta_i$ is the angle at which the crosslinked
filament connects to the test filament.

%tube potential k
%The tube potential is a convenient representation of many-body network
%effects on the test filament. However, the parameters entering the
%potential and their relation to network deformation and mechanical
%properties are usually unknown.  Ideally, these parameters can be
%determined in a self-consistent way from the analysis of the test
%filament.

We assume the network to be represented by an effective medium that
couples to the test filament only at the crosslinking points,
\begin{eqnarray}\label{eq:kx}
  k(s) = k_m\sum_{i=1}^{N}\alpha_i \delta(s-s_i)
\end{eqnarray}
where $k_m$ is the stiffness of the medium. $N$ is the total number of
crosslinking sites, and $\alpha_i=\sin^2\theta_i$ represents the
effects of the local network structure. The stiffness $k_m$ is thus
defined via the local network response to driving at a given crosslink
point. One may visualize this setting as a spring that is attached to
the polymer at the crosslink and that tries to force the polymer into
the strain-induced changing tube centerline. 

The central goal of this work is to calculate in a self-consistent way
the stiffness $k_m$, as well as its frequency-dependent
generalization, the complex modulus $g^\star(\omega)$. In previous
work we have argued that the stiffness may be calculated from the
equation
\begin{equation}\label{eq:self-consistence}
\frac{1}{2}k_m(\gamma L)^2 =\left\langle \min_{y(s)} \left(H_b[y]+V[y](k_m)\right)\right\rangle\,,
\end{equation}
where the angular brackets denote ensemble average with respect to the
quenched local network structure. This equation highlights the
two-fold role of the stiffness $k_m$. On a mesoscopic scale it is
defined as an elastic modulus that quantifies the energy cost to
deformation (left-hand side). On a microscopic scale this deformation
is carried by filaments that are themselves connected to the elastic
medium via the crosslinks (right-hand side).

Eq.~(\ref{eq:self-consistence}) can be solved in a simplified scaling
picture. To this end we assume one angle $\theta$, as well as one
wavelength $l_s\sim L/N$, to dominate. Minimization with respect to
$y$ then gives
\begin{equation}\label{eq:y_scaling}
y = \frac{k_m\sin^2\theta}{(\kappa/l_s^3)+k_m\sin^2\theta}\bar y
\end{equation}
Inserting in Eq.~(\ref{eq:self-consistence}) one finds
\begin{equation}\label{eq:}
1=N\cos^2\theta\frac{\kappa/l_s^3}{\kappa/l_s^3+k_m\sin^2\theta}
\end{equation}
which can be solved for the modulus $k$ 
\begin{equation}\label{eq:modulus_scaling_result}
k_m\sim \kappa N^3(N-N_c)
\end{equation}
where we defined $N_c=1/\cos^2\theta$, which represents the
percolation threshold of the model. The modulus is zero if less than
$N_c$ crosslinks are present, and scales with $\sim N^4$ far above the
threshold. We have shown previously~\cite{heussingerPRE2007} how the
inclusion of different wavelengths as well as angles can change the
scaling of the modulus with crosslink concentration $n$.

The static theory presented above has been used in various contexts,
e.g. to describe the mixing-rule in composite networks of microtubules
and f-actin~\cite{PhysRevLett.105.118101}. In the following we want to
generalize the theory to account for finite frequency of the
deformation, as well as for finite lifetime of the crosslink bond.

\section{Finite frequency}\label{sec:finite-frequency}

Experiments are most often conducted in the frequency domain, where a
time-dependent oscillatory strain $\gamma(t) = \gamma_0\sin(\omega t)$
is imposed. In order to account for time-dependent phenomena we first
rewrite Eq.~(\ref{eq:self-consistence}) as two force-balance
equations.

The minimization operation makes the transverse deflection of the
polymer, $y(s)$, the solution to the equation
\begin{eqnarray}\label{eq:bending_force_balance}
  0 = \kappa y^{(4)} + \sum_i\delta(s-s_i)T_i\sin\theta_i\,.
\end{eqnarray}
Here, we have defined the force in the ith crosslink
\begin{equation}\label{eq:axial_force_definition}
  T_i=k_m\sin\theta_i(y_i-\bar y_i)\,.
\end{equation}
The part of these forces transverse to the polymer ($T_i\sin\theta_i$)
must balance the bending force $\kappa y^{(4)}$ to give a stable
contour in mechanical equilibrium.

A second balance equation can be obtained by differentiating
Eq.~(\ref{eq:self-consistence}) with respect to $\gamma$. This will
give us the force that is needed to displace the polymer by the
strain. Using Eq.~(\ref{eq:self-consistence}) we find
\begin{equation}\label{eq:axial_force}
  k_m\gamma =
  \left\langle\sum_{i=1}^N T_i\cos\theta_i\right\rangle\,, 
\end{equation}
where now the forces $T_i$ are projected onto the axis of the
fiber. In other words, the external force $F_{\rm ext}=k_m\gamma$ is
balanced by the forces at the $n$ crosslinks.

The generalization to finite frequencies is now
straightforward. First, additional viscous (and possibly thermal)
forces need to enter the force-balance equations. Second, the
stiffness $k_m$ needs to be substituted by a frequency-dependent
function $g^\star(\omega)$. This is achieved by defining the response
function
\begin{equation}\label{eq:response_fct}
 T(t) = \int_{-\infty}^t d\tau
 g(t-\tau)\frac{\partial\gamma}{\partial\tau}\equiv (g\star \gamma)(t)\,.
\end{equation}
This function specifies the force at time $t$, that is needed for a
given strain history $\gamma(\tau)$.

If $g(t)=k_m$ is constant, then $T=k_m\gamma(t)$, i.e. a quasi-static
solid response, while the limit $g(t)=\zeta\delta(t)$ gives a
fluid-like behavior, where $T=\zeta\dot{\gamma}$.

With these modifications we obtain the following two equations:

\begin{eqnarray}\label{eq:transverse-forces-dynamic}
  \kappa y^{(4)} +
  \sum_{i}\alpha_i g\star(y-\bar y_i)\delta(s-s_i) \\\nonumber
= \eta\frac{\partial y}{\partial t} +\xi \,.
\end{eqnarray}

\begin{eqnarray}\label{eq:axial-forces-dynamics}
  g\star\gamma +
 \left\langle\sum_{i}\alpha_i\beta_ig\star(y-\bar
   y_i)\right\rangle\\\nonumber
=\eta_z\frac{\partial\gamma}{\partial t}+\xi_z \,.
\end{eqnarray}

Eq.(\ref{eq:transverse-forces-dynamic}) has to be solved for $y(s,t)$
and used in Eq.(\ref{eq:axial-forces-dynamics}) to determine the
response function $g(t)$, or in frequency-space
$g^\star(\omega)=g'(\omega) + ig''(\omega)$.

Adopting the latter representation,
Eq.~(\ref{eq:transverse-forces-dynamic}) can be written
as~\footnote{In the following we will neglect the axial damping
  $\eta_z$ as well as the noise terms $\xi$ and $\xi_z$. This does not
  qualitatively change the observed behavior.}
\begin{eqnarray}\label{eq:y_greens}
y(x,\omega) = \sum_i {\cal G}(x,x_i,\omega)g^\star\alpha_i(y_i-\bar y_i)
\end{eqnarray}
which shows how the Greens function ${\cal G}$ mediates between the
position $x_i$ of the crosslink, where the force $f =
g^\star\alpha_i(y_i-\bar y_i)$ is applied, and the actual position
$x$, at wich the deflection is evaluated. The Greens function itself
is given as
\begin{eqnarray}\label{eq:greens_fct}
{\cal G}_{ij}(\omega) \equiv{\cal G}(x_i,x_j,\omega) =
\sum_{q}\frac{\psi_q(x_i)\psi_q^\star(x_j)}{\kappa q^4+i\omega\eta}
\end{eqnarray}
where $\psi_q$ are suitable basis functions, e.g. trigonometric
functions that are chosen to respect the boundary conditions.

Inserting into Eq.~(\ref{eq:axial-forces-dynamics}) one obtains the
final equation
\begin{eqnarray}\label{eq:implicit_g}
1 = \left\langle\sum_{ij}(1+g^\star{\cal SG})^{-1}_{ij}\alpha_i\beta_i\beta_j\right\rangle
\end{eqnarray}
where we introduced the diagonal matrix $S_{ij}=\alpha_i\delta_{ij}$.
Eq.~(\ref{eq:implicit_g}) needs to be solved numerically for the
modulus $g^\star(\omega)$.

\begin{figure}
\centering
\includegraphics[width=0.38\linewidth]{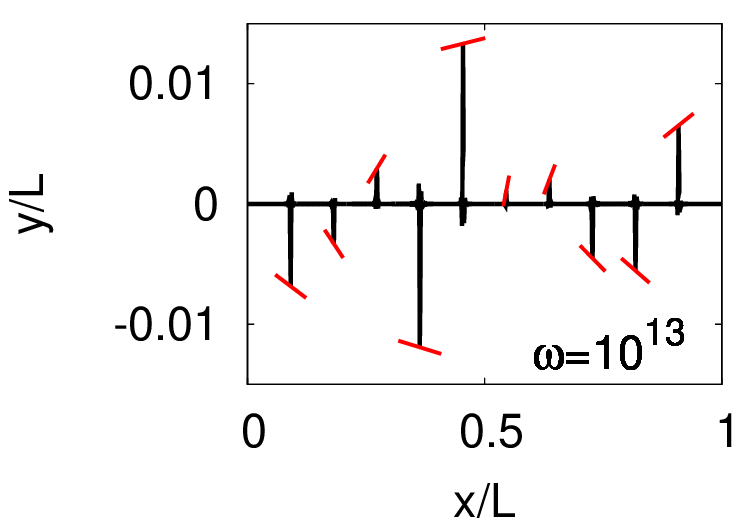}
\includegraphics[width=0.29\linewidth]{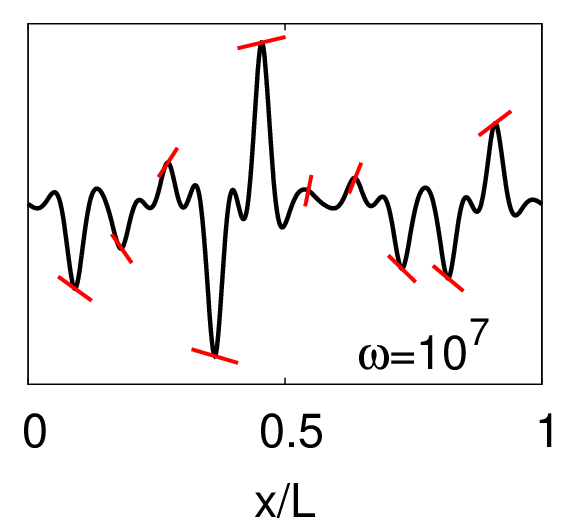}
\includegraphics[width=0.29\linewidth]{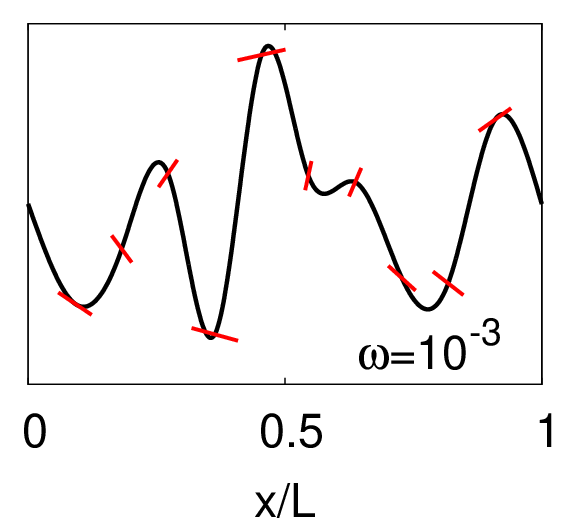}
\caption{\label{contour}Amplitude $y_0(\omega)$ of filament contour
  $y(x,t)=y_0(x)\sin(\omega t)$ for different driving frequencies. For
  high $\omega$ (left) the filament feels the driving only in the
  vicinity of the crosslinks (represented by the light (red)
  bars). Low-frequency driving only excites the longest possible
  wavelengths, which are set by the local network structure
  (orientation and distance of contacting filaments).}
\end{figure}

\begin{figure}
\centering
\includegraphics[width=0.9\linewidth]{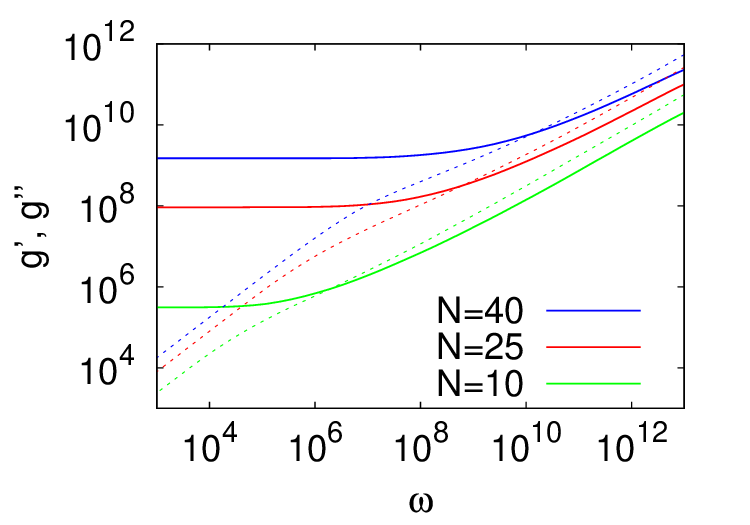}
\caption{\label{gstar_nobinding} Frequency-dependent modulus
  $g^\star=g'+ig''$ for different crosslink number $N$; storage
  modulus $g'$ -- solid lines; loss modulus $g''$ -- dashed lines.}
\end{figure}

%high frequencies
For high $\omega$ there is no coupling from one crosslink to the
next. The excited bending modes have small wavelength (see
Fig.~\ref{contour}), and perturbations are only local.
For high frequencies the Greens function is diagonal
\begin{eqnarray}\label{eq:green_high_omega}
%{\cal G}_{ij}(\omega) \to \delta_{ij}\frac{\sqrt{\pi}}{2(i\omega)^{3/4}}
{\cal G}_{ij}(\omega) \to \delta_{ij}\frac{1}{\sqrt{8}L\kappa^{1/4}(i\omega\eta)^{3/4}}
\end{eqnarray}
and the determining Eq.~(\ref{eq:implicit_g}) is simplified accordingly,
\begin{eqnarray}\label{eq:implicit_g_diagonal}
1 = \left\langle\sum_{i}\frac{\cos^2\theta_i}{1+g^\star\alpha_i{\cal G}_{ii}}\right\rangle
\end{eqnarray}
This gives $g^\star \sim (i\omega)^{3/4}$, as expected from the
dispersion ($\omega\sim q^{1/4}$) of the bending modes. The full
numerical solution of Eq.~(\ref{eq:implicit_g}) is presented in
Fig.~\ref{gstar_nobinding} for various crosslink densities $N$. On
small frequencies the static solution is recovered and leads to a
plateau in the storage modulus, $g_0\sim N^x$, where the value of the
exponent $x$ depends on the type of quenched local network structure
(angular brackets). The associated loss modulus scales linear with
frequency $g''\sim \omega$ and characterizes the viscous losses of a
filament that moves with a velocity $v\sim \omega$ through the
solvent.
%\xx{how does $g''$ scale with $N$ ? I guess linear ??}

\section{Finite crosslink lifetime}\label{sec:finite-crossl-lifet}

If thermal fluctuations are comparable to the strength of a crosslink,
then the bond will have a finite lifetime. In biological systems the
crosslink-induced bonds between filaments usually have a lifetime in
the range of seconds. The binding kinetics can therefore be picked up
in standard rheological measurements. In fact, some systems display a
pronounced peak in the loss modulus $g''$ at the respective
frequencies~\cite{PhysRevLett.101.108101,broederszPRL2010Linker}.

\begin{figure}[h]
  \centering
  \includegraphics[width=0.7\linewidth]{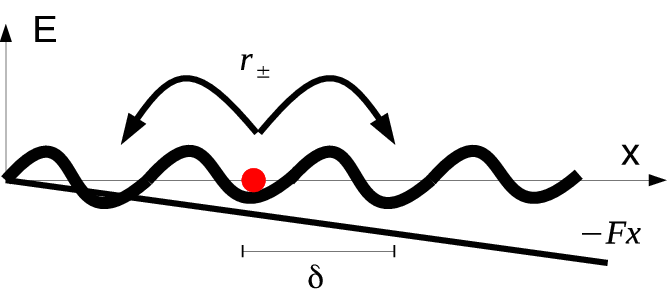}
  \caption{\label{crosslink_potential} Binding potential felt by the
    crosslink taken along the filament axis.  }
\end{figure}

In the following we explain how crosslink binding and unbinding can be
introduced into the theory. Alternative theoretical developments are
presented, for example, in
Refs.~\cite{C5SM00262A,PhysRevE.86.040901,heussinger12:_stres}. We
think of the crosslink to live in a one-dimensional periodic energy
landscape that represents the binding states along the filament
backbone (see Fig.~\ref{crosslink_potential}). In f-actin the
double-helical repeat implies a periodicity of roughly $\delta\approx
50$nm. While being bound at one site the crosslink stays in the
respective minimum of the energy landscape, unbinding corresponds to
Kramers escape from this minimum. In the fast-rebinding regime we can
assume the crosslink to immediately fall in the neighboring minimum a
distance $\delta$ away.

Via a force-dependent escape rate $r_\pm = r_0 e^{\pm \beta F\delta}$
one direction is favored over the other. In linear response the
crosslink then moves with a velocity $v=F/\zeta$ and friction
coefficient $\zeta=k_BT/r_o\delta^2$, as imposed by the fluctuation
dissipation relation and a diffusion constant $D=r_0\delta^2$.

We thus conclude that crosslink binding/rebinding processes can be
envisioned as a dash-pot that introduces viscous forces on the
filaments, the friction coefficient being given in terms of
microscopic properties of the crosslink and the binding domain of the
filament.

With this insight the response function $g$ on the right side of
Eqs.~(\ref{eq:transverse-forces-dynamic}) and
(\ref{eq:axial-forces-dynamics}) have to be substituted (in
frequency-space) by
\begin{eqnarray}\label{eq:serial_connection}
\bar g^{-1} = g^{-1} + (i\omega\zeta)^{-1}
\end{eqnarray}
representing a serial connection of crosslink binding domain $\zeta$
and visco-elastic medium $g$ (Maxwell element).

This modifies Eq.~(\ref{eq:implicit_g}) as follows
\begin{eqnarray}\label{eq:implicit_g_modified}
1 = \left\langle\sum_{ij}(1+g^\star\Lambda{\cal SG})^{-1}_{ij}\alpha_i\beta_i(\Lambda\beta)_j\right\rangle
\end{eqnarray}
the diagonal matrix $\Lambda$ containing the Maxwell elements of the
crosslink, $\Lambda_{ij}=\delta_{ij}\frac{i\omega\zeta_j}{i\omega\zeta_j+g}$.

%frequency-dependent modulus for different: ...

\begin{figure}
\centering
\includegraphics[width=0.9\linewidth]{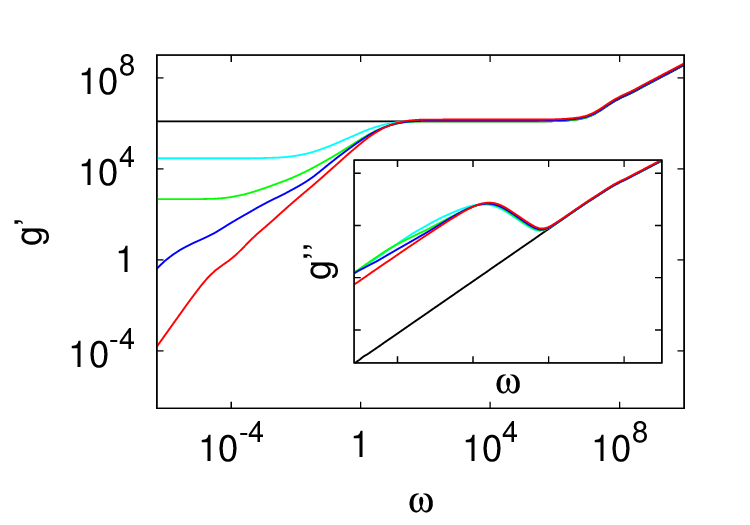}
\caption{\label{gstar_binding} Storage modulus (main panel) and loss
  modulus (inset, same axes as in main panel) vs. frequency for
  different fraction of quenched crosslinks $n_{\rm
    q}= N_{\rm q}/N=0,0.18,0.2,0.4,1$ (from bottom to top). A second plateau
  develops when the number of quenched crosslinks is above the
  percolation threhsold.}
\end{figure}

%mixture of quenched/binding crosslinks

%distribution of binding times

\begin{figure}
\centering
\includegraphics[width=0.9\linewidth]{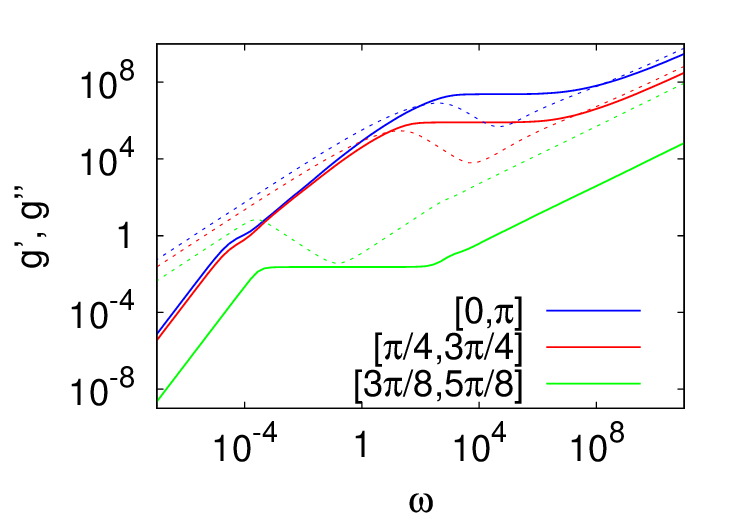}
\caption{\label{gstar_angular}Frequency-dependent modulus
  $g^\star=g'+ig''$ for different angular randomness $P(\theta)$ (flat
  distribution restricted to different intervals as specified in the
  legend). If crosslink intersection angles $\theta$ are sufficiently
  random, an anomalous regime at small frequencies develops that
  reflects the spatial inhomogeneity along the single filament.}
\end{figure}

The result of this calculation can be seen in
Figs.~\ref{gstar_binding} and \ref{gstar_angular}. Primarily,
crosslink binding leads to the appearance of a Maxwell-like peak at
small frequencies $\omega^\star\sim g_0/\zeta$, where $g_0$ is the
respective plateau modulus.

In Fig.~\ref{gstar_binding} we display the rheology for a mixture of
$N_{\rm r}$ reversible and $N_{\rm q}$ quenched (permanent)
crosslinks. If there is a minimum number of quenched crosslinks per
filament, there is a second plateau modulus at low frequencies. This
indicates that these quenched crosslinks are sufficient in number to
form a rigid structure -- rigidity percolates.

In Fig.~\ref{gstar_angular} we vary the structural randomness of the
network. In particular, the distribution $P(\theta)$ of crosslink
angles $\theta$ is changed. As a result, a broad intermediate regime
develops for the loss modulus, whenever the angles are broadly
distributed. This regime reflects the spatial heterogeneity along the
test filament. The ultimate low frequency regime ($g''\sim \omega^2$)
is only reached when all crosslinks along the test filament
effectively behave equally.

% relation to approach by Broedersz: there, anomalous tail
% $\omega^{1/2}$ described by Rouse-like relaxations. not included in
% present model.

\section{Nonlinear response}\label{sec:nonlinear-response}

A full nonlinear theory has to include several factors, e.g. the
reorientation of filaments under large
strain~\cite{1367-2630-17-8-083035} or the force-induced change in the
polymer end-to-end distance. Also the effects of an applied prestress
in combination with small amplitude oscillations is an important
experimental probe. It is outside the scope of this work to fully
combine all these aspects with our theoretical framework. However,
progress is possible on a ``schematic'' level.

\subsection{Prestress}

\begin{figure}[ht]
\centering
\includegraphics[width=0.8\linewidth]{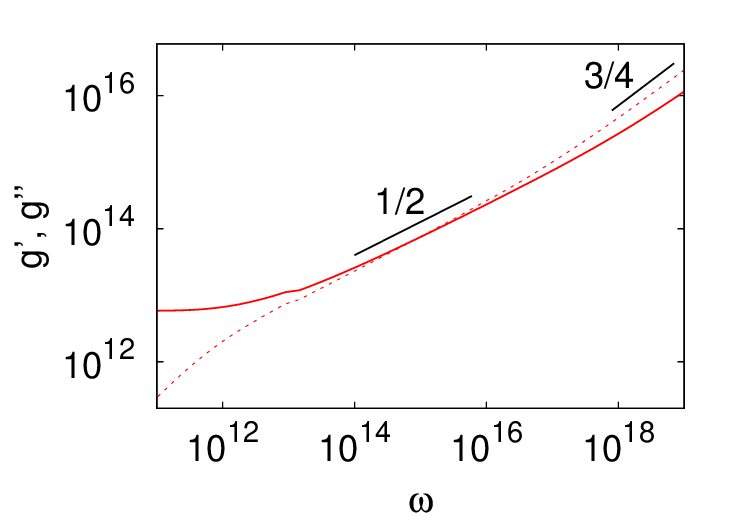}
\caption{\label{gstar_tension} Modulus vs frequency for a large value
  of tension ($f/g_0\approx10$).}
\end{figure}

To incorporate a constant prestress in our formalism, we make a
``quasi-linear'' approximation: We assume the linear theory to be
valid, while we change the propagator
\begin{equation} {\cal G}_{ij}(\omega) \to
  \sum_{q}\frac{\psi_q(x_i)\psi_q^\star(x_j)}{\kappa
    q^4+fq^2+i\omega\eta}
\end{equation}
where the new $f$-dependent term takes care of the reduction of
transverse undulations by applying a tensile prestress. This results
in a new stress-dependent plateau modulus $g_f\sim f$, as well as a
new regime $g\sim (i\omega)^{1/2}$ at intermediate frequency (see
Fig.~\ref{gstar_tension}). The frequency scale for this new scaling
regime is $\omega_f\sim f/\lambda^2\eta$, where $\lambda=L/N$ is the
wavelength of the relevant bending mode. In order for this regime to
be accessible, the tension needs to be large enough to make $g_f/g_0 >
(\omega_f/\omega_0)^{3/4}$, where $\omega_0\sim \kappa/\lambda^4\eta$
is the relevant frequency scale without tension.

%With our choice of units this implies $f>g_0^4$. ???? neglects
%prefactors ...

%In addition, there may be secondary effects of external tension: it
%may affect the distribution of crosslink lifetimes.

%We distinguish two effects: a shift of the average binding lifetime
%with force and a broadening of the distribution.

%see Fig.~\ref{gstar_lifetime}.

%\begin{figure}[ht]
%\centering
%%\includegraphics[width=0.8\linewidth]{pretension_modulus.eps}
%\caption{\label{gstar_lifetime} Modulus vs frequency for different
%  average binding lifetime and for changing width of the lifetime
%  distribution.}
%\end{figure}

%\begin{figure}[ht]
%\centering
%\includegraphics[width=0.8\linewidth]{pretension_modulus.eps}
%\caption{\label{gstar_binding} Modulus for different values of tension
%  $\tau$. For large tension and high frequencies a tension-dominated
%  regime $\sim\omega^{1/2}$ develops. \xx{unrealistic high values of
%    tensions ?1e9 in what units ?}. The plateau modulus increases as
%  ?? and the characteristic frequency shifts as ??.}
%\end{figure}

%what does prestress do on binding ?

%can prestress-dependent binding rates explain shift of maximum ? or
%can it also be a change in the distribution of binding rates ?

%binding-dependent prestress $f(\omega)$?

\subsection{Schematic theory for strain ramp}

Under large forces, the polymer will no longer behave as inextensible
rod. Rather the specific form of the force-extension relation will
become important. We can include this factor in a schematic model for
the behavior under a strain ramp~\cite{maier15}, where the strain
linearly increases in time, $\gamma(t) = \dot\gamma t$.

%The essential ingredient in our model of network viscoelasticity is
%that network deformation couples to nonaffine motion on the level of
%the filament such that bending modes 

%To set up such a description, we build on a model that
%has successfully been used to explain the linear rheology
%of bundled actin-fascin networks [31,35]. Related models
%have been proposed in [36,37]. 

%The general idea in our theoretical description is that, under large
%strain $\gamma$, two processes compete: first, non-linear filament
%elasticity beyond a strain Î³c leads to strain-stiffening; second,
%cross- link unbinding (at a rate koff ) leads to an increase in the
%wavelength Î» of the bending modes and subsequently to strain-softening
%[36,38]. It is this interplay between the strain-scale Î³c and the
%time-scale 1/koff that can generate a transition from stiffening to
%weakening as the strain rate Î³ Ì is varied (see fig. 3d).

This schematic model utilizes the key assumptions of
Sects.~\ref{sec:revi-stat-modul}-\ref{sec:finite-crossl-lifet}:
network strains translate into non-affine filament bending modes via a
deformation of the tube; the amplitude of these bends $\bar y \sim
\gamma L$ grows linearly with strain (see Eq.~(\ref{eq:bary})); the
wavelengths of the bends are slaved to the surrounding network
structure (factors $\beta(s)$; see Fig.~\ref{contour}). The bending
wavelengths are thus set by the typical inter-crosslink spacing. That
is, if we consider a filament with $N$ crosslinks, the average
bending wavelength will be $\lambda = L/N$.

Under larger strain, beyond the linear regime, two processes compete:
first, non-linear filament elasticity (nonlinear force-extension
relation) leads to strain-stiffening; second, cross- link unbinding
leads to an increase in the wavelength of the bending modes and
subsequently to strain-softening.% [36,38].

For a given bending amplitude $\bar y$, an associated longitudinal
extension $u$ (increase of end-to-end distance) can be calculated via
Pythagoras' law, $u \sim \bar y^2 /\lambda \sim \gamma^2 L^2/\lambda$.

In response to large elastic deformations the crosslinks start to unbind
%Now, we additionally assume that crosslinks can unbind such that the
%number of bound crosslinks n de- creases with time
(neglecting rebinding). Thus, the bending wavelength gets longer, as
$\lambda = L/N$, and the elastic energy decreases. The interplay
between stiffening and softening is then a competition between elastic
stiffening (embodied in the non-linear longitudinal response) and
softening via unbinding. To implement the softening part, we need a
model for the elastic energy as well as a dynamical evolution equation
for the crosslink number $N(t)$. The bending energy of the filament
scales $E_b = Nk_\perp \bar y^2 $ where we have used the bending
spring constant $k_\perp\sim \kappa/\lambda^3$ of an elastic filament
with bending stiffness $\kappa$.  For the stretching energy we take
the linearized force-extension relation of a wormlike chain with
spring constant $k_s \sim \kappa l_p/\lambda^4$ and the persistence
length $l_p$. The total energy then is $E = E_b + E_s$, the force $F$
is the first derivative, the modulus $\mu$ is the second derivative
with respect to strain. Without crosslink unbinding, this describes a
strain- stiffening system. The strain-dependence in the non-linear
regime follows from the longitudinal response and will be different,
for example, when one considers an exponential stiffening model as in
\cite{PhysRevE.86.040901}.

\begin{figure}[ht]
\centering
\includegraphics[width=0.45\linewidth]{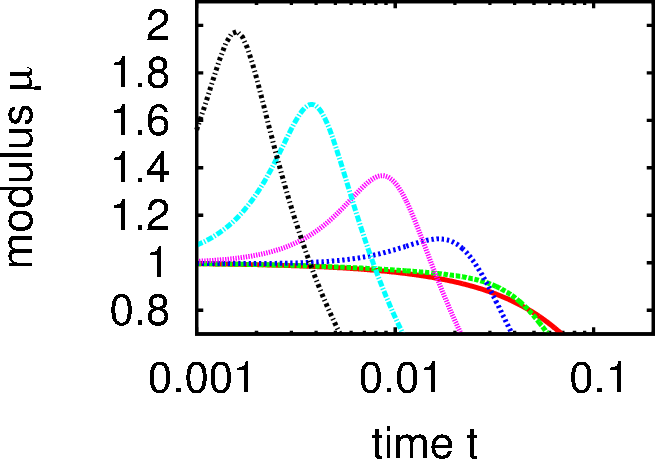}
\includegraphics[width=0.50\linewidth]{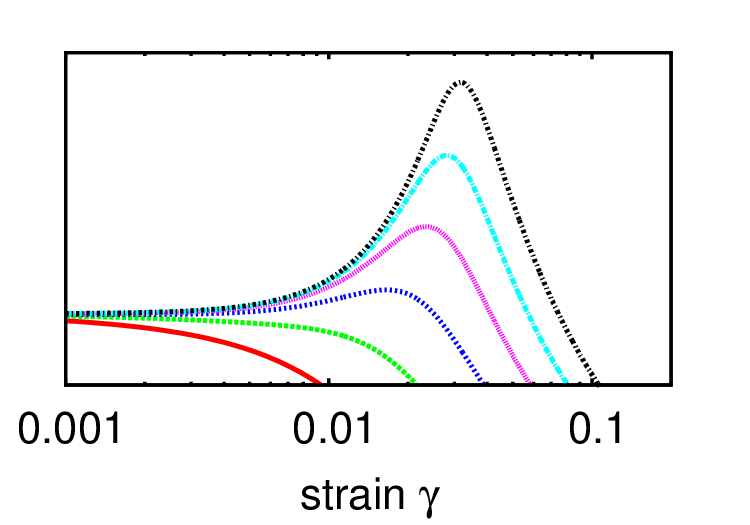}
\caption{\label{g_ramp} Nonlinear modulus $\mu$ vs time $t$ (left) and
  vs strain $\gamma$ (right) for different strain rates. Larger
  strainrates imply less time for unbinding processes. Therefore more
  of the elastic nonlinearity is visible in the modulus (the larger
  the strainrate the higher the peak).}\label{fig:strainramp}

\end{figure}

The simplest description for the crosslink dynamics is in terms of a
rate equation
\begin{equation}\label{eq:rate}
dN/dt = f(N) - b(N)\,,
\end{equation}
with forward rate $f$ and backward rate $b$. Neglecting rebinding,
$b = 0$. Unbinding happens at any one of $N$ crosslinks, thus
$f = N k_{\rm off}$, with an off-rate that may be force dependent,
$k_{\rm off} = k_0 e^{F(N)/F_0}$, with the $N$-dependent force $F(N)$
as given above.  Solving the combined problem then gives
Fig.~\ref{fig:strainramp}. Similar curves have been found
experimentally, for example in \cite{maier15,LielegSoftMatter2010,Semmrich18122007}.

\section{Conclusions}

In conclusion, we have presented a theoretical framework for the
linear and nonlinear visco-elastic properties of reversibly connected
networks of semiflexible polymers. In our model the network strain
does not couple directly to the filament end-to-end distance, but
rather serves to locally distort the network structure. This induces
bending modes in the filaments the amplitude of which grow linearly in
strain, and the wavelength of wich are slaved to the local network
structure, e.g. the distance to the next crosslink etc. Specifically,
we investigated the frequency-dependent linear rheology, in particular
in combination with crosslink binding/unbinding
processes. Furthermore, we devised a schematic model for the nonlinear
response in a creep experiment. These tests show that our model is
capable of reproducing many of the key experimental findings available
in the literature.

\section{Acknowledgments}

We acknowledge financial support by the German Science Foundation via
the Emmy Noether program (He 6322/1-1) and via the collaborative
research center SFB 937 (project A16).

%\bibliography{/home/claus/Documents/publ/papers/bib_files/en,/home/claus/Documents/publ/papers/bib_files/references.all}

\begin{thebibliography}{37}%
\makeatletter
\providecommand \@ifxundefined [1]{%
 \@ifx{#1\undefined}
}%
\providecommand \@ifnum [1]{%
 \ifnum #1\expandafter \@firstoftwo
 \else \expandafter \@secondoftwo
 \fi
}%
\providecommand \@ifx [1]{%
 \ifx #1\expandafter \@firstoftwo
 \else \expandafter \@secondoftwo
 \fi
}%
\providecommand \natexlab [1]{#1}%
\providecommand \enquote  [1]{``#1''}%
\providecommand \bibnamefont  [1]{#1}%
\providecommand \bibfnamefont [1]{#1}%
\providecommand \citenamefont [1]{#1}%
\providecommand \href@noop [0]{\@secondoftwo}%
\providecommand \href [0]{\begingroup \@sanitize@url \@href}%
\providecommand \@href[1]{\@@startlink{#1}\@@href}%
\providecommand \@@href[1]{\endgroup#1\@@endlink}%
\providecommand \@sanitize@url [0]{\catcode `\\12\catcode `\$12\catcode
  `\&12\catcode `\#12\catcode `\^12\catcode `\_12\catcode `\%12\relax}%
\providecommand \@@startlink[1]{}%
\providecommand \@@endlink[0]{}%
\providecommand \url  [0]{\begingroup\@sanitize@url \@url }%
\providecommand \@url [1]{\endgroup\@href {#1}{\urlprefix }}%
\providecommand \urlprefix  [0]{URL }%
\providecommand \Eprint [0]{\href }%
\providecommand \doibase [0]{http://dx.doi.org/}%
\providecommand \selectlanguage [0]{\@gobble}%
\providecommand \bibinfo  [0]{\@secondoftwo}%
\providecommand \bibfield  [0]{\@secondoftwo}%
\providecommand \translation [1]{[#1]}%
\providecommand \BibitemOpen [0]{}%
\providecommand \bibitemStop [0]{}%
\providecommand \bibitemNoStop [0]{.\EOS\space}%
\providecommand \EOS [0]{\spacefactor3000\relax}%
\providecommand \BibitemShut  [1]{\csname bibitem#1\endcsname}%
\let\auto@bib@innerbib\@empty
%</preamble>
\bibitem [{\citenamefont {Bausch}\ and\ \citenamefont {Kroy}(2006)}]{bausch06}%
  \BibitemOpen
  \bibfield  {author} {\bibinfo {author} {\bibfnamefont {A.}~\bibnamefont
  {Bausch}}\ and\ \bibinfo {author} {\bibfnamefont {K.}~\bibnamefont {Kroy}},\
  }\href@noop {} {\bibfield  {journal} {\bibinfo  {journal} {Nature Physics}\
  }\textbf {\bibinfo {volume} {2}},\ \bibinfo {pages} {231} (\bibinfo {year}
  {2006})}\BibitemShut {NoStop}%
\bibitem [{\citenamefont {Broedersz}\ and\ \citenamefont
  {MacKintosh}(2014)}]{RevModPhys.86.995}%
  \BibitemOpen
  \bibfield  {author} {\bibinfo {author} {\bibfnamefont {C.~P.}\ \bibnamefont
  {Broedersz}}\ and\ \bibinfo {author} {\bibfnamefont {F.~C.}\ \bibnamefont
  {MacKintosh}},\ }\href {\doibase 10.1103/RevModPhys.86.995} {\bibfield
  {journal} {\bibinfo  {journal} {Rev. Mod. Phys.}\ }\textbf {\bibinfo {volume}
  {86}},\ \bibinfo {pages} {995} (\bibinfo {year} {2014})}\BibitemShut
  {NoStop}%
\bibitem [{\citenamefont {Kroy}(2006)}]{kroy06:_elast}%
  \BibitemOpen
  \bibfield  {author} {\bibinfo {author} {\bibfnamefont {K.}~\bibnamefont
  {Kroy}},\ }\href@noop {} {\bibfield  {journal} {\bibinfo  {journal} {Curr.
  Op. Coll. Int. Sc.}\ }\textbf {\bibinfo {volume} {11}},\ \bibinfo {pages}
  {56} (\bibinfo {year} {2006})}\BibitemShut {NoStop}%
\bibitem [{\citenamefont {Gittes}\ and\ \citenamefont
  {MacKintosh}(1998)}]{gittesPRE1998}%
  \BibitemOpen
  \bibfield  {author} {\bibinfo {author} {\bibfnamefont {F.}~\bibnamefont
  {Gittes}}\ and\ \bibinfo {author} {\bibfnamefont {F.~C.}\ \bibnamefont
  {MacKintosh}},\ }\href@noop {} {\bibfield  {journal} {\bibinfo  {journal}
  {Phys. Rev. E}\ }\textbf {\bibinfo {volume} {58}},\ \bibinfo {pages} {R1241}
  (\bibinfo {year} {1998})}\BibitemShut {NoStop}%
\bibitem [{\citenamefont {Frey}\ \emph {et~al.}(1998)\citenamefont {Frey},
  \citenamefont {Kroy}, \citenamefont {Wilhelm},\ and\ \citenamefont
  {Sackmann}}]{frey98}%
  \BibitemOpen
  \bibfield  {author} {\bibinfo {author} {\bibfnamefont {E.}~\bibnamefont
  {Frey}}, \bibinfo {author} {\bibfnamefont {K.}~\bibnamefont {Kroy}}, \bibinfo
  {author} {\bibfnamefont {J.}~\bibnamefont {Wilhelm}}, \ and\ \bibinfo
  {author} {\bibfnamefont {E.}~\bibnamefont {Sackmann}},\ }\enquote {\bibinfo
  {title} {Dynamical networks in physics and biology},}\ \ (\bibinfo
  {publisher} {Springer},\ \bibinfo {address} {Berlin},\ \bibinfo {year}
  {1998})\ Chap.~\bibinfo {chapter} {9}\BibitemShut {NoStop}%
\bibitem [{\citenamefont {Morse}(2001)}]{PhysRevE.63.031502}%
  \BibitemOpen
  \bibfield  {author} {\bibinfo {author} {\bibfnamefont {D.~C.}\ \bibnamefont
  {Morse}},\ }\href {\doibase 10.1103/PhysRevE.63.031502} {\bibfield  {journal}
  {\bibinfo  {journal} {Phys. Rev. E}\ }\textbf {\bibinfo {volume} {63}},\
  \bibinfo {pages} {031502} (\bibinfo {year} {2001})}\BibitemShut {NoStop}%
\bibitem [{\citenamefont {Jones}\ and\ \citenamefont
  {Ball}(1991)}]{jones91:_elast_rigid_networ}%
  \BibitemOpen
  \bibfield  {author} {\bibinfo {author} {\bibfnamefont {J.~L.}\ \bibnamefont
  {Jones}}\ and\ \bibinfo {author} {\bibfnamefont {R.~C.}\ \bibnamefont
  {Ball}},\ }\href@noop {} {\bibfield  {journal} {\bibinfo  {journal}
  {Macromolecules}\ }\textbf {\bibinfo {volume} {24}},\ \bibinfo {pages} {6369}
  (\bibinfo {year} {1991})}\BibitemShut {NoStop}%
\bibitem [{\citenamefont {Wolff}\ \emph {et~al.}(2010)\citenamefont {Wolff},
  \citenamefont {Fernandez},\ and\ \citenamefont {Kroy}}]{WolffNJP2010}%
  \BibitemOpen
  \bibfield  {author} {\bibinfo {author} {\bibfnamefont {L.}~\bibnamefont
  {Wolff}}, \bibinfo {author} {\bibfnamefont {P.}~\bibnamefont {Fernandez}}, \
  and\ \bibinfo {author} {\bibfnamefont {K.}~\bibnamefont {Kroy}},\ }\href
  {http://stacks.iop.org/1367-2630/12/i=5/a=053024} {\bibfield  {journal}
  {\bibinfo  {journal} {New Journal of Physics}\ }\textbf {\bibinfo {volume}
  {12}},\ \bibinfo {pages} {053024} (\bibinfo {year} {2010})}\BibitemShut
  {NoStop}%
\bibitem [{\citenamefont {Heussinger}\ and\ \citenamefont
  {Frey}(2006)}]{heu06floppy}%
  \BibitemOpen
  \bibfield  {author} {\bibinfo {author} {\bibfnamefont {C.}~\bibnamefont
  {Heussinger}}\ and\ \bibinfo {author} {\bibfnamefont {E.}~\bibnamefont
  {Frey}},\ }\href@noop {} {\bibfield  {journal} {\bibinfo  {journal} {Phys.
  Rev. Lett.}\ }\textbf {\bibinfo {volume} {97}},\ \bibinfo {pages} {105501}
  (\bibinfo {year} {2006})}\BibitemShut {NoStop}%
\bibitem [{\citenamefont {Picu}(2011)}]{C1SM05022B}%
  \BibitemOpen
  \bibfield  {author} {\bibinfo {author} {\bibfnamefont {R.~C.}\ \bibnamefont
  {Picu}},\ }\href {\doibase 10.1039/C1SM05022B} {\bibfield  {journal}
  {\bibinfo  {journal} {Soft Matter}\ }\textbf {\bibinfo {volume} {7}},\
  \bibinfo {pages} {6768} (\bibinfo {year} {2011})}\BibitemShut {NoStop}%
\bibitem [{\citenamefont {Wilhelm}\ and\ \citenamefont
  {Frey}(2003)}]{Wilhelm2003}%
  \BibitemOpen
  \bibfield  {author} {\bibinfo {author} {\bibfnamefont {J.}~\bibnamefont
  {Wilhelm}}\ and\ \bibinfo {author} {\bibfnamefont {E.}~\bibnamefont {Frey}},\
  }\href@noop {} {\bibfield  {journal} {\bibinfo  {journal} {Physical Review
  Letters}\ }\textbf {\bibinfo {volume} {91(10)}} (\bibinfo {year}
  {2003})}\BibitemShut {NoStop}%
\bibitem [{\citenamefont {Head}\ \emph {et~al.}(2003)\citenamefont {Head},
  \citenamefont {Levine},\ and\ \citenamefont {MacKintosh}}]{Head2003}%
  \BibitemOpen
  \bibfield  {author} {\bibinfo {author} {\bibfnamefont {D.~A.}\ \bibnamefont
  {Head}}, \bibinfo {author} {\bibfnamefont {A.~J.}\ \bibnamefont {Levine}}, \
  and\ \bibinfo {author} {\bibfnamefont {F.~C.}\ \bibnamefont {MacKintosh}},\
  }\href@noop {} {\bibfield  {journal} {\bibinfo  {journal} {Physical Review
  Letters}\ }\textbf {\bibinfo {volume} {91(10)}} (\bibinfo {year}
  {2003})}\BibitemShut {NoStop}%
\bibitem [{\citenamefont {{\AA}strom}\ \emph {et~al.}(2009)\citenamefont
  {{\AA}strom}, \citenamefont {Kumar},\ and\ \citenamefont
  {Karttunen}}]{B815892D}%
  \BibitemOpen
  \bibfield  {author} {\bibinfo {author} {\bibfnamefont {J.~A.}\ \bibnamefont
  {{\AA}strom}}, \bibinfo {author} {\bibfnamefont {P.~B.~S.}\ \bibnamefont
  {Kumar}}, \ and\ \bibinfo {author} {\bibfnamefont {M.}~\bibnamefont
  {Karttunen}},\ }\href {\doibase 10.1039/B815892D} {\bibfield  {journal}
  {\bibinfo  {journal} {Soft Matter}\ }\textbf {\bibinfo {volume} {5}},\
  \bibinfo {pages} {2869} (\bibinfo {year} {2009})}\BibitemShut {NoStop}%
\bibitem [{\citenamefont {M\"uller}\ and\ \citenamefont
  {Kierfeld}(2014)}]{PhysRevLett.112.094303}%
  \BibitemOpen
  \bibfield  {author} {\bibinfo {author} {\bibfnamefont {P.}~\bibnamefont
  {M\"uller}}\ and\ \bibinfo {author} {\bibfnamefont {J.}~\bibnamefont
  {Kierfeld}},\ }\href {\doibase 10.1103/PhysRevLett.112.094303} {\bibfield
  {journal} {\bibinfo  {journal} {Phys. Rev. Lett.}\ }\textbf {\bibinfo
  {volume} {112}},\ \bibinfo {pages} {094303} (\bibinfo {year}
  {2014})}\BibitemShut {NoStop}%
\bibitem [{\citenamefont {Bai}\ \emph {et~al.}(2011)\citenamefont {Bai},
  \citenamefont {Missel}, \citenamefont {Klug},\ and\ \citenamefont
  {Levine}}]{C0SM00496K}%
  \BibitemOpen
  \bibfield  {author} {\bibinfo {author} {\bibfnamefont {M.}~\bibnamefont
  {Bai}}, \bibinfo {author} {\bibfnamefont {A.~R.}\ \bibnamefont {Missel}},
  \bibinfo {author} {\bibfnamefont {W.~S.}\ \bibnamefont {Klug}}, \ and\
  \bibinfo {author} {\bibfnamefont {A.~J.}\ \bibnamefont {Levine}},\ }\href
  {\doibase 10.1039/C0SM00496K} {\bibfield  {journal} {\bibinfo  {journal}
  {Soft Matter}\ }\textbf {\bibinfo {volume} {7}},\ \bibinfo {pages} {907}
  (\bibinfo {year} {2011})}\BibitemShut {NoStop}%
\bibitem [{\citenamefont {MacKintosh}\ \emph {et~al.}(1995)\citenamefont
  {MacKintosh}, \citenamefont {K{\"a}s},\ and\ \citenamefont {Janmey}}]{mac95}%
  \BibitemOpen
  \bibfield  {author} {\bibinfo {author} {\bibfnamefont {F.~C.}\ \bibnamefont
  {MacKintosh}}, \bibinfo {author} {\bibfnamefont {J.}~\bibnamefont {K{\"a}s}},
  \ and\ \bibinfo {author} {\bibfnamefont {P.~A.}\ \bibnamefont {Janmey}},\
  }\href@noop {} {\bibfield  {journal} {\bibinfo  {journal} {Phys. Rev. Lett.}\
  }\textbf {\bibinfo {volume} {75}},\ \bibinfo {pages} {4425} (\bibinfo {year}
  {1995})}\BibitemShut {NoStop}%
\bibitem [{\citenamefont {Storm}\ \emph {et~al.}(2005)\citenamefont {Storm},
  \citenamefont {Pastore}, \citenamefont {MacKintosh}, \citenamefont
  {Lubensky},\ and\ \citenamefont {Janmey}}]{Storm2005}%
  \BibitemOpen
  \bibfield  {author} {\bibinfo {author} {\bibfnamefont {C.}~\bibnamefont
  {Storm}}, \bibinfo {author} {\bibfnamefont {J.~J.}\ \bibnamefont {Pastore}},
  \bibinfo {author} {\bibfnamefont {F.~C.}\ \bibnamefont {MacKintosh}},
  \bibinfo {author} {\bibfnamefont {T.~C.}\ \bibnamefont {Lubensky}}, \ and\
  \bibinfo {author} {\bibfnamefont {P.~A.}\ \bibnamefont {Janmey}},\
  }\href@noop {} {\bibfield  {journal} {\bibinfo  {journal} {Nature}\ }\textbf
  {\bibinfo {volume} {435}},\ \bibinfo {pages} {191} (\bibinfo {year}
  {2005})}\BibitemShut {NoStop}%
\bibitem [{\citenamefont {Kroy}\ and\ \citenamefont
  {Glaser}(2007)}]{kroyNJP2007}%
  \BibitemOpen
  \bibfield  {author} {\bibinfo {author} {\bibfnamefont {K.}~\bibnamefont
  {Kroy}}\ and\ \bibinfo {author} {\bibfnamefont {J.}~\bibnamefont {Glaser}},\
  }\href@noop {} {\bibfield  {journal} {\bibinfo  {journal} {New J. Phys.}\
  }\textbf {\bibinfo {volume} {9}},\ \bibinfo {pages} {416} (\bibinfo {year}
  {2007})}\BibitemShut {NoStop}%
\bibitem [{\citenamefont {Das}\ \emph {et~al.}(2012)\citenamefont {Das},
  \citenamefont {Quint},\ and\ \citenamefont
  {Schwarz}}]{das12:_redun_cooper_mechan_compos_cross_filam_networ}%
  \BibitemOpen
  \bibfield  {author} {\bibinfo {author} {\bibfnamefont {M.}~\bibnamefont
  {Das}}, \bibinfo {author} {\bibfnamefont {D.~A.}\ \bibnamefont {Quint}}, \
  and\ \bibinfo {author} {\bibfnamefont {J.~M.}\ \bibnamefont {Schwarz}},\
  }\href@noop {} {\bibfield  {journal} {\bibinfo  {journal} {PLoS One}\
  }\textbf {\bibinfo {volume} {7}},\ \bibinfo {pages} {35939} (\bibinfo {year}
  {2012})}\BibitemShut {NoStop}%
\bibitem [{\citenamefont {Mao}\ \emph {et~al.}(2013)\citenamefont {Mao},
  \citenamefont {Stenull},\ and\ \citenamefont
  {Lubensky}}]{PhysRevE.87.042601}%
  \BibitemOpen
  \bibfield  {author} {\bibinfo {author} {\bibfnamefont {X.}~\bibnamefont
  {Mao}}, \bibinfo {author} {\bibfnamefont {O.}~\bibnamefont {Stenull}}, \ and\
  \bibinfo {author} {\bibfnamefont {T.~C.}\ \bibnamefont {Lubensky}},\ }\href
  {\doibase 10.1103/PhysRevE.87.042601} {\bibfield  {journal} {\bibinfo
  {journal} {Phys. Rev. E}\ }\textbf {\bibinfo {volume} {87}},\ \bibinfo
  {pages} {042601} (\bibinfo {year} {2013})}\BibitemShut {NoStop}%
\bibitem [{\citenamefont {Sheinman}\ \emph {et~al.}(2012)\citenamefont
  {Sheinman}, \citenamefont {Broedersz},\ and\ \citenamefont
  {MacKintosh}}]{PhysRevE.85.021801}%
  \BibitemOpen
  \bibfield  {author} {\bibinfo {author} {\bibfnamefont {M.}~\bibnamefont
  {Sheinman}}, \bibinfo {author} {\bibfnamefont {C.~P.}\ \bibnamefont
  {Broedersz}}, \ and\ \bibinfo {author} {\bibfnamefont {F.~C.}\ \bibnamefont
  {MacKintosh}},\ }\href {\doibase 10.1103/PhysRevE.85.021801} {\bibfield
  {journal} {\bibinfo  {journal} {Phys. Rev. E}\ }\textbf {\bibinfo {volume}
  {85}},\ \bibinfo {pages} {021801} (\bibinfo {year} {2012})}\BibitemShut
  {NoStop}%
\bibitem [{\citenamefont {Broedersz}\ \emph {et~al.}(2011)\citenamefont
  {Broedersz}, \citenamefont {Mao}, \citenamefont {Lubensky},\ and\
  \citenamefont {MacKintosh}}]{c.11:_critic_isost}%
  \BibitemOpen
  \bibfield  {author} {\bibinfo {author} {\bibfnamefont {C.~P.}\ \bibnamefont
  {Broedersz}}, \bibinfo {author} {\bibfnamefont {X.}~\bibnamefont {Mao}},
  \bibinfo {author} {\bibfnamefont {T.~C.}\ \bibnamefont {Lubensky}}, \ and\
  \bibinfo {author} {\bibfnamefont {F.~C.}\ \bibnamefont {MacKintosh}},\
  }\href@noop {} {\bibfield  {journal} {\bibinfo  {journal} {Nat. Phys.}\
  }\textbf {\bibinfo {volume} {7}},\ \bibinfo {pages} {983} (\bibinfo {year}
  {2011})}\BibitemShut {NoStop}%
\bibitem [{\citenamefont {Liu}\ \emph {et~al.}(2010)\citenamefont {Liu},
  \citenamefont {Nagel}, \citenamefont {van Saarloos},\ and\ \citenamefont
  {M.Wyart}}]{LiuNagelSaarloosWyartREVIEW2010}%
  \BibitemOpen
  \bibfield  {author} {\bibinfo {author} {\bibfnamefont {A.~J.}\ \bibnamefont
  {Liu}}, \bibinfo {author} {\bibfnamefont {S.~R.}\ \bibnamefont {Nagel}},
  \bibinfo {author} {\bibfnamefont {W.}~\bibnamefont {van Saarloos}}, \ and\
  \bibinfo {author} {\bibnamefont {M.Wyart}},\ }\enquote {\bibinfo {title} {The
  jamming scenario -- an introduction and outlook},}\ \ (\bibinfo  {publisher}
  {Oxford University Press},\ \bibinfo {year} {2010})\ Chap.~\bibinfo {chapter}
  {9}\BibitemShut {NoStop}%
\bibitem [{\citenamefont {Heussinger}\ \emph {et~al.}(2007)\citenamefont
  {Heussinger}, \citenamefont {Schaefer},\ and\ \citenamefont
  {Frey}}]{heussingerPRE2007}%
  \BibitemOpen
  \bibfield  {author} {\bibinfo {author} {\bibfnamefont {C.}~\bibnamefont
  {Heussinger}}, \bibinfo {author} {\bibfnamefont {B.}~\bibnamefont
  {Schaefer}}, \ and\ \bibinfo {author} {\bibfnamefont {E.}~\bibnamefont
  {Frey}},\ }\href@noop {} {\bibfield  {journal} {\bibinfo  {journal} {Phys.
  Rev. E}\ }\textbf {\bibinfo {volume} {76}},\ \bibinfo {pages} {031906}
  (\bibinfo {year} {2007})}\BibitemShut {NoStop}%
\bibitem [{\citenamefont {Heussinger}\ and\ \citenamefont
  {Frey}(2007)}]{heussingerEPJE2007}%
  \BibitemOpen
  \bibfield  {author} {\bibinfo {author} {\bibfnamefont {C.}~\bibnamefont
  {Heussinger}}\ and\ \bibinfo {author} {\bibfnamefont {E.}~\bibnamefont
  {Frey}},\ }\href@noop {} {\bibfield  {journal} {\bibinfo  {journal} {Eur.
  Phys. J. E}\ }\textbf {\bibinfo {volume} {24}},\ \bibinfo {pages} {47}
  (\bibinfo {year} {2007})}\BibitemShut {NoStop}%
\bibitem [{\citenamefont {Broedersz}\ \emph {et~al.}(2012)\citenamefont
  {Broedersz}, \citenamefont {Sheinman},\ and\ \citenamefont
  {MacKintosh}}]{PhysRevLett.108.078102}%
  \BibitemOpen
  \bibfield  {author} {\bibinfo {author} {\bibfnamefont {C.~P.}\ \bibnamefont
  {Broedersz}}, \bibinfo {author} {\bibfnamefont {M.}~\bibnamefont {Sheinman}},
  \ and\ \bibinfo {author} {\bibfnamefont {F.~C.}\ \bibnamefont {MacKintosh}},\
  }\href {\doibase 10.1103/PhysRevLett.108.078102} {\bibfield  {journal}
  {\bibinfo  {journal} {Phys. Rev. Lett.}\ }\textbf {\bibinfo {volume} {108}},\
  \bibinfo {pages} {078102} (\bibinfo {year} {2012})}\BibitemShut {NoStop}%
\bibitem [{\citenamefont {Huisman}\ \emph {et~al.}(2010)\citenamefont
  {Huisman}, \citenamefont {Heussinger}, \citenamefont {Storm},\ and\
  \citenamefont {Barkema}}]{PhysRevLett.105.118101}%
  \BibitemOpen
  \bibfield  {author} {\bibinfo {author} {\bibfnamefont {E.~M.}\ \bibnamefont
  {Huisman}}, \bibinfo {author} {\bibfnamefont {C.}~\bibnamefont {Heussinger}},
  \bibinfo {author} {\bibfnamefont {C.}~\bibnamefont {Storm}}, \ and\ \bibinfo
  {author} {\bibfnamefont {G.~T.}\ \bibnamefont {Barkema}},\ }\href {\doibase
  10.1103/PhysRevLett.105.118101} {\bibfield  {journal} {\bibinfo  {journal}
  {Phys. Rev. Lett.}\ }\textbf {\bibinfo {volume} {105}},\ \bibinfo {pages}
  {118101} (\bibinfo {year} {2010})}\BibitemShut {NoStop}%
\bibitem [{Note1()}]{Note1}%
  \BibitemOpen
  \bibinfo {note} {In the following we will neglect the axial damping $\eta _z$
  as well as the noise terms $\xi $ and $\xi _z$. This does not qualitatively
  change the observed behavior.}\BibitemShut {Stop}%
\bibitem [{\citenamefont {Lieleg}\ \emph {et~al.}(2008)\citenamefont {Lieleg},
  \citenamefont {Claessens}, \citenamefont {Luan},\ and\ \citenamefont
  {Bausch}}]{PhysRevLett.101.108101}%
  \BibitemOpen
  \bibfield  {author} {\bibinfo {author} {\bibfnamefont {O.}~\bibnamefont
  {Lieleg}}, \bibinfo {author} {\bibfnamefont {M.~M. A.~E.}\ \bibnamefont
  {Claessens}}, \bibinfo {author} {\bibfnamefont {Y.}~\bibnamefont {Luan}}, \
  and\ \bibinfo {author} {\bibfnamefont {A.~R.}\ \bibnamefont {Bausch}},\
  }\href {\doibase 10.1103/PhysRevLett.101.108101} {\bibfield  {journal}
  {\bibinfo  {journal} {Phys. Rev. Lett.}\ }\textbf {\bibinfo {volume} {101}},\
  \bibinfo {pages} {108101} (\bibinfo {year} {2008})}\BibitemShut {NoStop}%
\bibitem [{\citenamefont {Broedersz}\ \emph {et~al.}(2010)\citenamefont
  {Broedersz}, \citenamefont {Depken}, \citenamefont {Yao}, \citenamefont
  {Pollak}, \citenamefont {Weitz},\ and\ \citenamefont
  {MacKintosh}}]{broederszPRL2010Linker}%
  \BibitemOpen
  \bibfield  {author} {\bibinfo {author} {\bibfnamefont {C.~P.}\ \bibnamefont
  {Broedersz}}, \bibinfo {author} {\bibfnamefont {M.}~\bibnamefont {Depken}},
  \bibinfo {author} {\bibfnamefont {N.~Y.}\ \bibnamefont {Yao}}, \bibinfo
  {author} {\bibfnamefont {M.~R.}\ \bibnamefont {Pollak}}, \bibinfo {author}
  {\bibfnamefont {D.~A.}\ \bibnamefont {Weitz}}, \ and\ \bibinfo {author}
  {\bibfnamefont {F.~C.}\ \bibnamefont {MacKintosh}},\ }\href {\doibase
  10.1103/PhysRevLett.105.238101} {\bibfield  {journal} {\bibinfo  {journal}
  {Phys. Rev. Lett.}\ }\textbf {\bibinfo {volume} {105}},\ \bibinfo {pages}
  {238101} (\bibinfo {year} {2010})}\BibitemShut {NoStop}%
\bibitem [{\citenamefont {Vaca}\ \emph {et~al.}(2015)\citenamefont {Vaca},
  \citenamefont {Shlomovitz}, \citenamefont {Yang}, \citenamefont {Valentine},\
  and\ \citenamefont {Levine}}]{C5SM00262A}%
  \BibitemOpen
  \bibfield  {author} {\bibinfo {author} {\bibfnamefont {C.}~\bibnamefont
  {Vaca}}, \bibinfo {author} {\bibfnamefont {R.}~\bibnamefont {Shlomovitz}},
  \bibinfo {author} {\bibfnamefont {Y.}~\bibnamefont {Yang}}, \bibinfo {author}
  {\bibfnamefont {M.~T.}\ \bibnamefont {Valentine}}, \ and\ \bibinfo {author}
  {\bibfnamefont {A.~J.}\ \bibnamefont {Levine}},\ }\href {\doibase
  10.1039/C5SM00262A} {\bibfield  {journal} {\bibinfo  {journal} {Soft Matter}\
  }\textbf {\bibinfo {volume} {11}},\ \bibinfo {pages} {4899} (\bibinfo {year}
  {2015})}\BibitemShut {NoStop}%
\bibitem [{\citenamefont {Wolff}\ and\ \citenamefont
  {Kroy}(2012)}]{PhysRevE.86.040901}%
  \BibitemOpen
  \bibfield  {author} {\bibinfo {author} {\bibfnamefont {L.}~\bibnamefont
  {Wolff}}\ and\ \bibinfo {author} {\bibfnamefont {K.}~\bibnamefont {Kroy}},\
  }\href {\doibase 10.1103/PhysRevE.86.040901} {\bibfield  {journal} {\bibinfo
  {journal} {Phys. Rev. E}\ }\textbf {\bibinfo {volume} {86}},\ \bibinfo
  {pages} {040901} (\bibinfo {year} {2012})}\BibitemShut {NoStop}%
\bibitem [{\citenamefont {Heussinger}(2012)}]{heussinger12:_stres}%
  \BibitemOpen
  \bibfield  {author} {\bibinfo {author} {\bibfnamefont {C.}~\bibnamefont
  {Heussinger}},\ }\href@noop {} {\bibfield  {journal} {\bibinfo  {journal}
  {New J. Phys.}\ }\textbf {\bibinfo {volume} {14}},\ \bibinfo {pages} {095029}
  (\bibinfo {year} {2012})}\BibitemShut {NoStop}%
\bibitem [{\citenamefont {Amuasi}\ \emph {et~al.}(2015)\citenamefont {Amuasi},
  \citenamefont {Heussinger}, \citenamefont {Vink},\ and\ \citenamefont
  {Zippelius}}]{1367-2630-17-8-083035}%
  \BibitemOpen
  \bibfield  {author} {\bibinfo {author} {\bibfnamefont {H.~E.}\ \bibnamefont
  {Amuasi}}, \bibinfo {author} {\bibfnamefont {C.}~\bibnamefont {Heussinger}},
  \bibinfo {author} {\bibfnamefont {R.~L.~C.}\ \bibnamefont {Vink}}, \ and\
  \bibinfo {author} {\bibfnamefont {A.}~\bibnamefont {Zippelius}},\ }\href
  {http://stacks.iop.org/1367-2630/17/i=8/a=083035} {\bibfield  {journal}
  {\bibinfo  {journal} {New Journal of Physics}\ }\textbf {\bibinfo {volume}
  {17}},\ \bibinfo {pages} {083035} (\bibinfo {year} {2015})}\BibitemShut
  {NoStop}%
\bibitem [{\citenamefont {Maier}\ \emph {et~al.}(2015)\citenamefont {Maier},
  \citenamefont {Müller}, \citenamefont {Heussinger}, \citenamefont {Köhler},
  \citenamefont {Wall}, \citenamefont {Bausch},\ and\ \citenamefont
  {Lieleg}}]{maier15}%
  \BibitemOpen
  \bibfield  {author} {\bibinfo {author} {\bibfnamefont {M.}~\bibnamefont
  {Maier}}, \bibinfo {author} {\bibfnamefont {K.}~\bibnamefont {Müller}},
  \bibinfo {author} {\bibfnamefont {C.}~\bibnamefont {Heussinger}}, \bibinfo
  {author} {\bibfnamefont {S.}~\bibnamefont {Köhler}}, \bibinfo {author}
  {\bibfnamefont {W.}~\bibnamefont {Wall}}, \bibinfo {author} {\bibfnamefont
  {A.}~\bibnamefont {Bausch}}, \ and\ \bibinfo {author} {\bibfnamefont
  {O.}~\bibnamefont {Lieleg}},\ }\href {\doibase 10.1140/epje/i2015-15050-3}
  {\bibfield  {journal} {\bibinfo  {journal} {The European Physical Journal E}\
  }\textbf {\bibinfo {volume} {38}},\ \bibinfo {eid} {50} (\bibinfo {year}
  {2015}),\ 10.1140/epje/i2015-15050-3}\BibitemShut {NoStop}%
\bibitem [{\citenamefont {Lieleg}\ \emph {et~al.}(2010)\citenamefont {Lieleg},
  \citenamefont {Claessens},\ and\ \citenamefont
  {Bausch}}]{LielegSoftMatter2010}%
  \BibitemOpen
  \bibfield  {author} {\bibinfo {author} {\bibfnamefont {O.}~\bibnamefont
  {Lieleg}}, \bibinfo {author} {\bibfnamefont {M.~M. A.~E.}\ \bibnamefont
  {Claessens}}, \ and\ \bibinfo {author} {\bibfnamefont {A.~R.}\ \bibnamefont
  {Bausch}},\ }\href@noop {} {\bibfield  {journal} {\bibinfo  {journal} {Soft
  Matter}\ }\textbf {\bibinfo {volume} {6}},\ \bibinfo {pages} {218} (\bibinfo
  {year} {2010})}\BibitemShut {NoStop}%
\bibitem [{\citenamefont {Semmrich}\ \emph {et~al.}(2007)\citenamefont
  {Semmrich}, \citenamefont {Storz}, \citenamefont {Glaser}, \citenamefont
  {Merkel}, \citenamefont {Bausch},\ and\ \citenamefont
  {Kroy}}]{Semmrich18122007}%
  \BibitemOpen
  \bibfield  {author} {\bibinfo {author} {\bibfnamefont {C.}~\bibnamefont
  {Semmrich}}, \bibinfo {author} {\bibfnamefont {T.}~\bibnamefont {Storz}},
  \bibinfo {author} {\bibfnamefont {J.}~\bibnamefont {Glaser}}, \bibinfo
  {author} {\bibfnamefont {R.}~\bibnamefont {Merkel}}, \bibinfo {author}
  {\bibfnamefont {A.~R.}\ \bibnamefont {Bausch}}, \ and\ \bibinfo {author}
  {\bibfnamefont {K.}~\bibnamefont {Kroy}},\ }\href {\doibase
  10.1073/pnas.0705513104} {\bibfield  {journal} {\bibinfo  {journal}
  {Proceedings of the National Academy of Sciences}\ }\textbf {\bibinfo
  {volume} {104}},\ \bibinfo {pages} {20199} (\bibinfo {year} {2007})},\
  \Eprint
  {http://arxiv.org/abs/http://www.pnas.org/content/104/51/20199.full.pdf+html}
  {http://www.pnas.org/content/104/51/20199.full.pdf+html} \BibitemShut
  {NoStop}%
\end{thebibliography}


\end{document}